# Quantification of 2D Interfaces: Quality of heterostructures, and what is inside a nanobubble

*Mainak Mondal[1], Pawni Manchanda[2], Soumadeep Saha[1], Abhishek Jangid[1], Akshay Singh[1] ***

[1] Department of Physics, Indian Institute of Science, Bengaluru, India- 560012

[2] Department of Physical Sciences, Indian Institute of Science Education and Research (IISER) Kolkata, Mohanpur, India- 741246

KEYWORDS: two-dimensional materials, heterostructures, interface quality, nanobubble, optical contrast, optical microscopy

ABSTRACT: Trapped materials at the interfaces of two-dimensional heterostructures (HS) lead to reduced coupling between the layers, resulting in degraded optoelectronic performance and device variability. Further, nanobubbles can form at the interface during transfer or after annealing. The question of what is inside a nanobubble, i.e. the trapped material, remains unanswered, limiting the studies and applications of these nanobubble systems. In this work, we report two key advances. Firstly, we quantify the interface quality using RAW-format optical imaging, and distinguish between ideal and non-ideal interfaces. The HS-substrate ratio value is calculated using a transfer matrix model, and is able to detect the presence of trapped layers. The second key advance is identification of water as the trapped material inside a nanobubble. To the best of our knowledge, this is the first study to show that optical imaging alone can quantify interface quality, and find the type of trapped material inside spontaneously formed nanobubbles. We also define a quality index parameter to quantify the interface quality of HS. Quantitative measurement of the interface will help answer the question whether annealing is necessary during HS preparation, and will enable creation of complex HS with small twist angles. Identification of the trapped materials will pave the way towards using nanobubbles for novel optical and engineering applications.

**Introduction:**

The optoelectronic properties of two-dimensional (2D) materials are highly tunable by applied strain[1,2], temperature[3], and electric[4] fields, making 2D materials highly versatile for next-generation optoelectronics and quantum technologies[5]. The versatility of 2D materials increases many-fold by stacking different (or the same) 2D materials on top of one another, creating an entirely new class of structures called 2D heterostructures (HS)[6]. These HS are playgrounds for both fundamental physics problems[7], as well as for device applications[8]. HS are commonly prepared by stacking 2D layers using polymer-based pickup and drop methods, broadly called dry transfer techniques[9]. During the transfer process, some materials (polymer, water, air) can get trapped at the interface of individual layers, resulting in higher interlayer distance (higher than van der Waals (vdW) distance)[10]. Increased interlayer distance reduces the interlayer coupling between layers[11] and degrades the device performance[12]. Hence, proper evaluation of interface quality is of utmost importance in understanding experimental observations, and for optimization of devices.

Annealing of the HS at high temperatures in an inert or vacuum environment is often used to improve the interface quality, but in most cases creates pockets of trapped materials and nanobubbles[13,14]. Studies have shown that the pressure inside these nanobubbles can vary from tens of MPa to few GPa[15]. Such extremely localized pressure provides opportunities to locally engineer the HS's properties for various applications such as efficient single-photon sources[16], enhanced photodetection[17], localized doping[18], and bandgap tuning[19]. The mechanics of these pockets are primarily governed by the adhesion between vdW layers and properties of the trapped material[13,15,20]. However, the chemical composition of these trapped materials is usually unknown, leading to limited optimization possibility of transfer processes and device variability[19].

Photoluminescence (PL)[21], Raman spectroscopy[21,22], atomic force microscopy (AFM)[23], and electrical measurements[12] (depending on the HS) are conventionally used to determine the interface quality, although being time-consuming and expensive. On the other hand, optical microscope imaging-based methods can be rapid and easy-to-implement for large-scale characterization, especially when scalability is a major concern. Although several studies show that optical imaging can be used to identify single or few layers of 2D materials[24,25], it is seldom used to quantitatively determine the interface quality of HS. Our previous study showed that for conventional RGB (red-green-blue) format imaging, the contrast between the substrate and 2D materials is non-universal due to the associated post-processing in digital image processing[26]. Thus, RGB-format imaging cannot be used to quantitively characterize the interface of HS. On the other hand, RAW-format imaging is a linear intensity format, can quantify layer number[27,28], and amenable to contrast calculations via transfer matrix model for various combinations of 2D materials and substrates[26]. Hence, RAW-format imaging and the transfer matrix model calculation can provide a robust yet easy-to-implement technique to study systems beyond the complexity of individual 2D materials.

In this work, we studied two types of HS: $MoS_2/MoS_2$ (homo-HS) and $MoSe_2/WSe_2$ (hetero-HS). The $MoS_2/MoS_2$ HS consists of natural trilayers (N-TL) adjacent to a stacked trilayer (S-TL). Using RAW-format imaging, we have compared the layer-substrate ratio values of the S-TL region with the N-TL region before and after annealing. Before annealing, the stacked region's ratio values differ significantly from the natural regions, particularly for the blue channel, which means that the interface is not ideal due to trapped materials. After annealing, separate regions are formed inside the stacked region with different ratio values. We determine that the regions with the same (different) ratio values as natural regions have near-ideal (non-ideal) interfaces. PL mapping on the HS confirmed the interface quality and agreed with our RAW-format image analysis conclusions. The same technique is then used for $MoSe_2/WSe_2$ bilayer HS (BL-HS) as well, for which a natural HS does not exist. Further, we propose using a quality index parameter to quantify the interface quality of a specific HS region. The ability to quantify the quality of the interface before annealing is expected to be a game-changer for device fabrication, especially for twisted HS and complex architectures which are modified by annealing[7]. Then, to understand the nanobubbles formed after annealing, we measured the spatial profile of HS using AFM and calculated the ratio values considering trapped materials (air, water, and PDMS) in the HS. We find that the calculation for water as trapped material matched excellently with the measurement, thus identifying the material trapped inside. The identification of trapped materials inside nanobubbles will help us understand the optical properties of these nanobubbles. This work is expected to greatly help the 2D material community in optimizing the HS fabrication process by quantifying the interface quality.

# Results and Discussion:

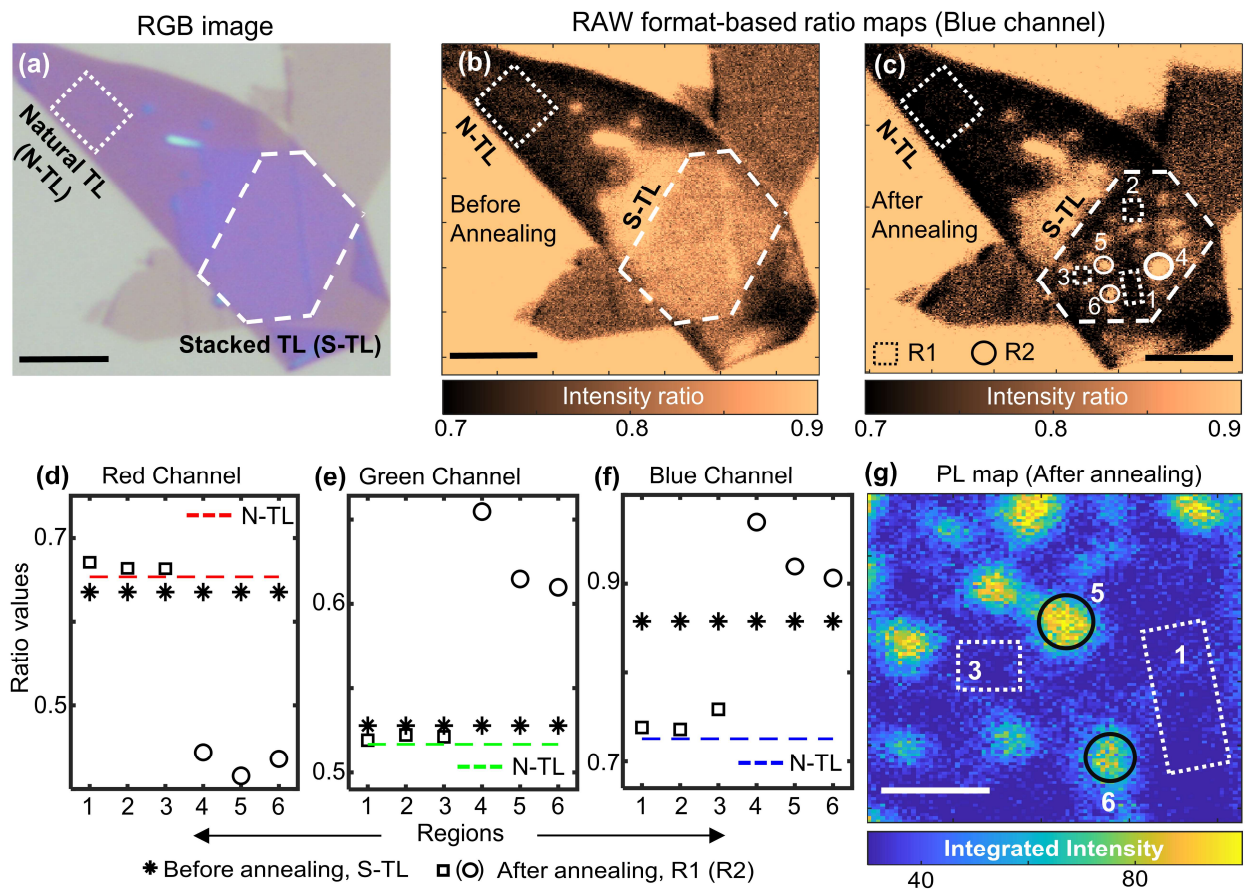

**Figure 1 (Determining the interface quality of a MoS$_2$ stacked trilayer by analyzing the RAW-format based intensity ratio values).** (a) RGB-format image of a MoS$_2$ heterostructure (HS) on a 257 nm SiO$_2$/Si substrate. Dashed lines on the image mark natural trilayer (N-TL) and stacked trilayer (S-TL) regions. (b) and (c) RAW-format based blue channel intensity ratio maps (each pixel's intensity values divided by the mean value of substrate region) of the HS before and after annealing. In Figure (b), dashed lines mark the N-TL and S-TL regions. In Figure (c), regions appearing similar to (different from) N-TL are marked with rectangular (circular) apertures and named R1 (R2). (d), (e), and (f) Variation of ratio values from before and after annealing for different regions are shown corresponding to blue, green, and red channels, respectively. The dashed line represents the ratio values from the N-TL region; star signs represent the values of the S-TL region before annealing. Rectangular (circular) markers represent the ratio values of rectangular (circular) aperture regions, marked in Figure (c). (g) Zoomed in integrated Photoluminescence (PL) intensity map of the HS after annealing to show the regions 1, 3, 5, and 6. R1 and R2 regions are marked with rectangular and circular apertures inside the S-TL region, the same as Figure (c). The scale bar in Figure (a), (b), and (c) is 5 μm. The scale bar in Figure (g) is 1 μm. All the measurements are performed at room temperature.

First, we exfoliated $MoS_2$ using the conventional scotch tape method. Then, a $MoS_2$ flake, consisting of a natural trilayer (N-TL) attached with a natural bilayer (BL), was transferred onto an isolated monolayer (ML) flake (kept on a 257 nm $SiO_2$/Si substrate) using the PDMS method (see Methods for more details of sample preparation). Specifically, the BL is placed on the ML to form the stacked TL (S-TL). RGB-format optical image of this HS is shown in Figure 1(a). N-TL and S-TL regions are marked on the image with dashed lines. This sample architecture allows us to simultaneously image both natural and stacked TL regions, i.e., compare an ideal interface with a stacked interface. We can thus investigate the interface quality of the stacked region, and *quantify* the interface quality.

As evident from the RGB-format optical image (Figure 1(a)), although the S-TL region appears to be a uniform region, the color does not visually match the N-TL region. We have used RAW-format imaging-based ratio mapping to further understand the difference between these two regions. In ratio mapping, the intensity values of each pixel are divided by the mean intensity value of the substrate region. The ratio map corresponding to the blue channel is shown in Figure 1(b), and green and red channel ratio maps are shown in Supporting Information, Section 2. A significant difference is observed in the ratio values of N-TL and S-TL. This difference in ratio values can be explained by considering a layer of trapped material in the stacked region. The presence of this trapped material significantly changes the reflectance of the stacked region; as a result, the apparent ratio values are much different from those of the N-TL region, which has no trapped material inside (Supporting Information, Section 1). Thus, the experimentalist can take an informed decision whether to anneal, and improve coupling, by just measuring the RAW-format image.

Next, we annealed the HS by heating it at 150 ºC for 15 minutes, a conventional method to improve the interface quality[29]. The RGB-format image of the annealed HS is shown in Supporting Information, Section 2. RAW-format blue channel ratio map of the after-annealing HS is shown in Figure 1(c), green and red channel ratio maps are shown in Supporting Information, Section 2. Upon annealing, a drastic change is observed in the stacked region. Some of the regions' ratio values are the same as the N-TL region's ratio value and are marked with the rectangular boxes in Figure 1(c), and named 1, 2, and 3 (R1 category). On the other hand, some of the regions' ratio values moved further away from the N-TL region's ratio value and are marked with circles in Figure 1(c), and named 4, 5, and 6 (R2 category).

To visualize the evolution of the interface after annealing, we plot the mean values of these regions 1 to 6 in Figure 1(f). Rectangular markers represent R1 category regions, and circular markers represent R2 category regions. Star-shaped markers represent the mean values of the before-annealing S-TL. The dashed line represents the mean value of the N-TL (ideal HS). Similarly, ratio values for green and red channels (region-wise) are plotted in Figure 1 (d) and (e), respectively. We observed that regions from the R1 category matched with N-TL or approached the N-TL ratio value for all the channels. From these observations, we conclude that the R1 category regions have no detectable trapped material, and have ideal interface quality and coupling between the layers. For R2 category regions, the ratio values moved much further from the N-TL regions for all the channels, even further than the pre-annealing S-TL. AFM measurements on the after-annealed HS shows that the R2 category regions are now turned into nanobubbles (> 30 nm height), which were not present before annealing (AFM images are shown in Supporting Information, Section 3). The formation of bubbles confirms that the quantity of trapped material has now increased inside the R2 category regions, resulting in much different ratio values. Hence, the interface quality of R2 regions is worse than R1. To further confirm the interface quality, we have measured the integrated PL map; the zoomed-in image of the map is shown in Figure 1 (g), and the complete map is in Supporting Information, Section 4. Similar to Figure 1(c), R1 (R2) category regions are marked with rectangular (circular) regions in Figure 1 (g). In the integrated PL intensity map, R1 category regions show much less PL than the R2 category regions, meaning that the R1 regions are well coupled and behave like TL (giving significantly less PL)[21]. Meanwhile, for R2 regions, the bottom ML does not couple well with the BL because of a significant amount of trapped material inside, resulting in higher PL emission than R1[21]. Thus, the interface quality evaluation using the RAW-format channel-based intensity ratio analysis agrees with the conventional PL measurement.

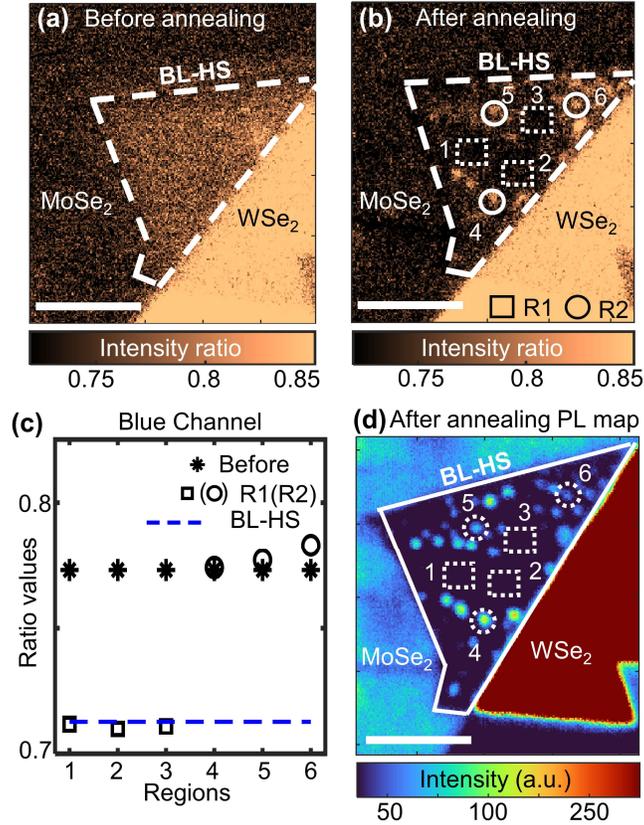

**Figure 2 (Determining the interface quality of a MoSe$_2$/WSe$_2$ bilayer heterostructure (BL-HS) by analyzing the RAW-format based intensity ratio values).** (a) and (b) RAW format-based blue channel ratio map of the MoSe$_2$/WSe$_2$ BL-HS before and after annealing. In Figures (a) and (b), BL-HS region is marked with white boundaries, and individual layered regions are specified. In Figure (b), regions where the ratio values have decreased (increased or remain the same) after annealing are marked with rectangular (circular) apertures and categorized as R1 (R2). (c) Variations of ratio values from before and after annealing for different regions correspond to the blue channel. The dashed line represents the calculated ratio value for an ideal MoSe$_2$/WSe$_2$ BL-HS; star signs represent the values of the BL-HS region before annealing. Rectangular (circular) markers represent the ratio values of rectangular (circular) aperture regions, marked in Figure (b). (d) Integrated Photoluminescence (PL) intensity map of the HS after annealing, R1 and R2 regions are marked with rectangular and circular apertures inside the BL-HS region. All scale bars are 3 μm. All measurements are performed at room temperature.

To showcase the robustness of our technique, we applied the ratio analysis technique on a $MoSe_2/WSe_2$ BL-HS, for which a natural analogue does not exist. We have exfoliated $MoSe_2$ and $WSe_2$ MLs from bulk crystals and transferred the $WSe_2$ ML on top of the $MoSe_2$ ML to create the BL-HS, using the same dry transfer method as discussed above. Unlike the $MoS_2/MoS_2$ HS discussed in Figure 1, hetero BL-HS (such as $MoSe_2/WSe_2$) cannot have a natural stacked region for direct comparison. Hence, we have calculated the ratio value of an ideal $MoSe_2/WSe_2$ BL-HS using transfer matrix model calculations[26,30,31]. After the transfer (before annealing), the BL-HS region appeared to be uniform (RGB-format optical image of this device is shown in Supporting Information, Section 5), which is also reflected in the RAW-format blue channel ratio map, shown in Figure 2(a). However, upon annealing, a drastic change is observed in BL-HS ratio values, which is shown in Figure 2(b). Some regions' ratio values decreased and became similar to the calculated ratio value of the ideal BL-HS (regions marked with rectangles), whereas other regions' values remained the same or slightly increased (regions marked with circles). RGB-format images and RAW-format ratio maps of before and after annealing BL-HS are shown in Supporting Information, Section 5.

To visualize these changes further, we plotted the variation of ratio values from before and after annealing in Figure 2(c). For regions 1 to 3 (R1 category, rectangular markers), ratio values have decreased, and for regions 4 to 6 (R2 category, circular markers), ratio values remained the same or slightly increased from the before annealing values (star-shaped markers). The calculated ratio value of the ideal BL-HS is shown as a dashed blue line in Figure 2(c). R1's ratio values match the calculated ideal BL-HS's values, whereas R2's values are further apart. Therefore, we conclude that R1 has an ideal-like interface quality, whereas R2's interface quality is non-ideal.

To further support our conclusions, we have measured the integrated PL intensity map of this BL-HS, shown in Figure 2(d) (different regions are marked in the same manner as Figure 2(b)). Type-II band alignment is formed by coupling $MoSe_2$ and $WSe_2$ MLs, wherein the generated electrons (holes) of the $WSe_2$ ($MoSe_2$) layer transfer to the $MoSe_2$ ($WSe_2$) layer and form interlayer excitons[32]. These interlayer excitons mostly recombine non-radiatively at room temperature and quench intralayer exciton emission, resulting in poor luminescence from properly coupled regions compared to individual layers[33]. Hence, the PL map confirms that the R1 category regions (appearing dark) are ideally coupled, whereas the R2 regions (appearing bright) are non-ideally coupled. The PL mapping thus matches our conclusions from RAW-ratio analysis. Thus, we proved that our RAW-format ratio analysis method is robust enough to evaluate interface quality for both homo and heterostructures. We also note that whereas the PL technique may only be applied to certain HS, the proposed RAW-format contrast technique is applicable to all HS.

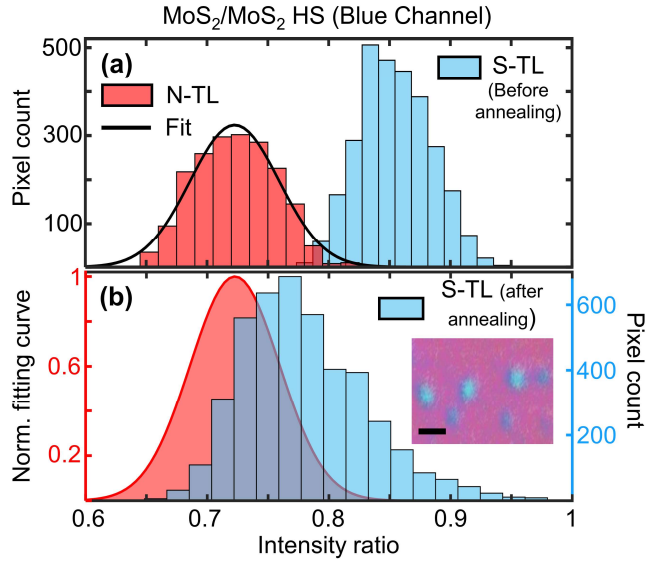

**Figure 3 (Quality index for the interface quality). (a)** Histogram showing the distribution of pixel counts with ratio values for N-TL and S-TL before annealing. Gaussian fit of the N-TL distribution is shown in a solid black line. **(b)** The ratio values distribution for the after-annealing S-TL region (shown in the inset) is plotted on the right y-axis. The y-axis left shows the normalized fitted curve for the N-TL region. This data is for the $MoS_2/MoS_2$ device, blue channel (discussed in Figure 1). The scale bar of the inset is 1 μm.

By comparing the average ratio values of different regions, we have successfully qualitatively determined the near-ideal and non-ideal quality of homo/heterostructure interfaces and confirmed those using conventional PL measurements. Now, we develop a quantitative parameter to quantify the interface quality. First, we use the N-TL region ($MoS_2/MoS_2$ HS discussed in Figure 1), which acts as the reference for quality quantification, and extract the distribution of the number of pixels with ratio values (Figure 3(a)). This distribution is fitted with a Gaussian, shown in a black solid line in Figure 3(a), with the width of the distribution attributed to signal to noise in the CMOS image sensor. Signal to noise can be increased (or width can be decreased) with increasing illumination. We are working in the regime of high signal to noise; a detailed discussion is given in Supporting Information, Section 6. Similarly, the distribution of the S-TL region, before and after annealing, is shown in Figures 3(a) and 3(b), respectively. The ratio map for the region of HS used to measure the distribution (for S-TL) is shown as an inset in Figure 3(b). Before and after annealing distributions (for S-TL) reveal the significant shift of ratio values towards the N-TL region's distribution, which matches the observation discussed in Figure 1(f). We also note that the S-TL ratio distribution has become broader after annealing, but has at least some overlap with the N-TL distribution, signifying the partially beneficial impact of annealing.

We now introduce a quality index parameter to quantify the stacked region's interface quality. The quality index value is defined as 1 (0) when the distribution of ratio values for N-TL and stacked region is properly overlapped (separated). We have normalized the N-TL region's fitted distribution to calculate the quality index and used this fitted distribution as a weight function. Next, we multiplied this weight function with the S-TL region's distribution to quantify the overlap of the N-TL distribution with the S-TL distribution. This distribution overlap, normalized with the total S-TL distribution's area, is defined as the quality index. The quality index of the region shown in Figure 3(b) is 0.45. We also note that for the $MoS_2/MoS_2$ HS on 257 nm $SiO_2$/Si substrate, our calculations show that blue and red channel ratio values should change nearly equally for the trapped material (air, water, or PDMS) thickness of less than 10 nm (shown in Supporting Information, Section 1). However, we observe that the blue channel ratio changes are higher than in red channel (which needs more understanding). Hence, we consider the blue channel to quantify the quality index of this HS (quality index for other channels are shown in Supporting Information, Section 7). For different HS configurations, this choice of color channel must be considered.

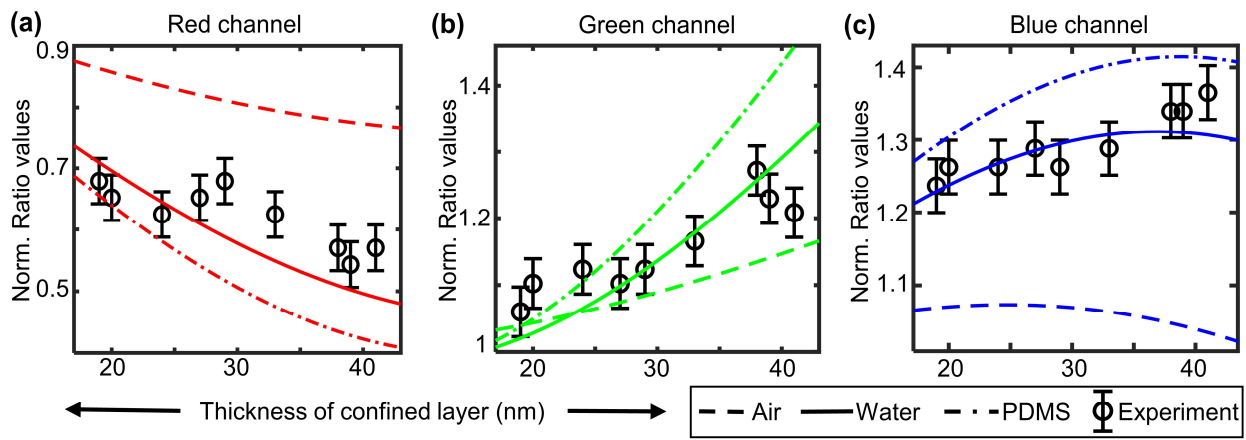

**Figure 4 (Comparison of experimentally measured and calculated ratio values from nanobubbles to determine the trapped material inside).** (a), (b), and (c) Experimentally measured ratio values are compared with the calculated ratio values for red, green, and blue channels. Error bars are the standard deviation of the ratio value measurement. Each ratio value is normalized with the ratio value of natural trilayers (N-TL) of $MoS_2$. Calculated ratio values (for red, green, and blue channels) of the S-TL region for different trapped layer thicknesses are shown as dashed, solid, and dash-dot lines for air, water, and PDMS as trapped material, respectively.

As discussed in Figure 1, a uniform interface is observed at the stacked region before annealing because of the uniform distribution of the trapped layer at the interface. Upon annealing, the trapped material is redistributed and accumulates in certain areas, forming a bubble structure (R2 in Figure 1(b)), but with unknown composition. In other words, we do not know what is inside the nanobubble. We propose using a combination of RAW-format imaging and AFM imaging to determine the refractive index of the trapped material. First, we measure the topography of these bubbles using AFM measurements to get direct height information (Supporting Information, Section 3). Next, the transfer matrix model is used to calculate the reflectance of this HS for varying thickness of the trapped layer[26]. We have considered air (n = 1), water (n = 1.33), and PDMS (n = 1.40)[34] as the possible trapped layers, since these will be present in the transfer process[10,35,36]. The intensity values are then normalized with the ideal interface case (i.e., zero nm trapped layer) and plotted for all channels in Figure 4. The dash-dot and solid lines represent the calculated ratio values for PDMS and water, respectively. It is evident from the calculations that with varying thickness of trapped material, the ratio values can change significantly. Also, different trapped materials can cause experimentally distinguishable variations in ratio values for the same thickness, and thus the trapped materials can be identified. We note that curvature of the nanobubbles has not been considered in the calculations.

Next, we compared the experimentally measured ratio values with the calculations. We consider a 10×10-pixel region (in the optical image) centered on the bubble, and selected the extremum of ratio value (depending on the channel) corresponding to that region's highest thickness. For the blue (red) channel, the highest (lowest) ratio value is selected, keeping the trend of ratio values with increasing thickness in mind (see Figure 4). Then, we extract the maximum height of a bubble from the AFM measurement, by applying a moving average over 400 nm × 400 nm areas on the AFM map. This averaging is required because AFM's spatial resolution is much higher than the optical microscope's resolution (constrained by the optical diffraction limit). We have plotted the normalized ratio values with corresponding trapped layer thickness (= measured height of the bubble – the height at the clean interface) for red, green, and blue channels in Figure 4(a), (b), and (c). The error bar represents the standard deviation of the pixel's intensity value distribution for uniform N-TL region. Interestingly, when considering water as the trapped layer, the experimentally found ratio values closely match the calculations. Hence, using this RAW-format imaging-based technique (coupled with AFM), we confirm that water is spontaneously trapped while transferring, which several other researchers have previously suggested, but never confirmed[13,35,37].

**Conclusion:**

In this study, we have made two key advances. The first key advance is that we have quantified the interface quality of 2D-HS, by using RAW format-based ratio (HS-substrate) values via an optical microscope. To demonstrate this process, we prepared a $MoS_2/MoS_2$ HS with natural and stacked trilayer regions, and imaged this HS before- and after-annealing. Before annealing, the ratio value map of the S-TL region shows significant differences from the N-TL region, confirming the presence of a uniform trapped layer between the stacked layers. After annealing, ratio values for some regions of the S-TL match the N-TL, whereas other regions become much different. These comparisons led to a direct confirmation of the interface quality of a particular region, which is then confirmed by conventional PL mapping. The same technique is used for $MoSe_2/WSe_2$ BL-HS to verify the interface quality, for which a natural HS does not exist. Further, we propose a quality index parameter to quantify the interface quality of the stacked region, by comparing the ratio value distribution of stacked regions with natural regions. The second key advance is the identification of the trapped materials inside a nanobubble. To understand the nanobubbles created at a 2D-HS interface, we have measured the ratio values for multiple nanobubbles with different thicknesses. Then, we compared the calculated ratio values with the AFM measurements, and identified water as the trapped material.

Through this study, we have solved the long-standing need for a simple but robust characterization technique to quantify the interface quality of 2D-HS. This method can easily be extended to any interface, including 2D materials and thin films. This work will be especially useful to answer the question whether annealing is necessary to create a high-quality HS, and will help experimentalists to create complex HS with small twist angles and various 2D layers. Identification of trapped materials inside the nanobubbles is a crucial step towards modifying these nanobubbles for various applications.

**Methods:**

Sample preparation: $MoS_2$, $WSe_2$ and $MoSe_2$ bulk crystals are purchased from 2D Semiconductors. Bulk crystals are exfoliated using conventional scotch tape method. The scotch is placed on a PDMS sheet to get the few layer flakes, which are then confirmed using RAW-format ratio values[26]. Flakes are then selectively transferred from the PDMS sheet to the 257 nm $SiO_2$/Si substrate at 60 ⁰C temperature using a custom-built transfer (micro-manipulator) setup.

Optical measurements: All shown optical images are taken using the Olympus BX53M microscope equipped with a 100X (0.9 numerical aperture) objective, white LED and Amscope CMOS MU1803. Integrated PL map is measured using Picoquant Micro Time 200 setup. In this setup, 507 nm laser is focused on the sample using 100X (0.9 numerical aperture) objective and the emission is passed through a 600 nm long pass filter and measured by an Excelitas single photon avalanche diode.

Atomic Force Microscopy: Park NX20 is used in non-contact mode to perform the atomic force microscopy on the samples.

ASSOCIATED CONTENT

**Supporting Information**

Section 1: Transfer matrix model and ratio values calculations; Section 2: RGB optical image and RAW format-based ratio maps for $MoS_2/MoS_2$ HS; Section 3: AFM maps of $MoS_2/MoS_2$ HS before and after annealing; Section 4: Photoluminescence map of $MoS_2/MoS_2$ HS after annealing; Section 5: RGB optical image and RAW format-based ratio maps for $MoSe_2/WSe_2$

BL-HS; Section 6: Variation of ratio values distribution width with increasing illumination; Section 7: Quality index calculations for $MoS_2/MoS_2$ HS.

AUTHOR INFORMATION

**Corresponding Author**

* Akshay Singh, aksy@iisc.ac.in

**Author Contributions**

MM and AS developed the experimental and theoretical framework. MM, PM, SS performed the sample preparation and optical experiments. MM, PM performed the data analysis. AJ performed the AFM measurements. MM and AS discussed and prepared the manuscript, with contributions from PM, SS and AJ.

**Data Availability**

All data is available upon reasonable request.

ACKNOWLEDGMENT

AS would like to ac knowledge funding from Indian Institute of Science start-up grant and DST Nanomission CONCEPT (Consortium for Collective and Engineered Phenomena in Topology) grant. MM would like to acknowledge Prime Minister's Research Fellowship (PMRF). SS would like to acknowledge Kishore Vaigyanik Protsahana Yojana (KVPY) for scholarship. The authors also acknowledge Micro Nano Characterization Facility (MNCF), Centre for Nano Science and Engineering (CeNSE), for the use of characterization facilities.

ABBREVIATIONS

RGB, red-green-blue; AFM, atomic force microscope; PL, photoluminescence; ML, monolayer; BL, bilayer; TL, trilayer; N-TL, natural trilayer; S-TL, stacked trilayer; HS, heterostructure; 2D, two-dimensional; vdW, van der Waals.

REFERENCES

(1) Conley, H. J.; Wang, B.; Ziegler, J. I.; Haglund, R. F. Jr.; Pantelides, S. T.; Bolotin, K. I. Bandgap Engineering of Strained Monolayer and Bilayer MoS2. *Nano Lett.* **2013**, *13* (8), 3626–3630.

(2) Datye, I. M.; Daus, A.; Grady, R. W.; Brenner, K.; Vaziri, S.; Pop, E. Strain-Enhanced Mobility of Monolayer MoS2. *Nano Lett.* **2022**, *22* (20), 8052–8059.

(3) Liu, H.-L.; Yang, T.; Chen, J.-H.; Chen, H.-W.; Guo, H.; Saito, R.; Li, M.-Y.; Li, L.-J. Temperature-Dependent Optical Constants of Monolayer MoS2, MoSe2, WS2, and WSe2: Spectroscopic Ellipsometry and First-Principles Calculations. *Sci Rep* **2020**, *10* (1), 15282.

(4) Ross, J. S.; Wu, S.; Yu, H.; Ghimire, N. J.; Jones, A. M.; Aivazian, G.; Yan, J.; Mandrus, D. G.; Xiao, D.; Yao, W.; Xu, X. Electrical Control of Neutral and Charged Excitons in a Monolayer Semiconductor. *Nature Communications* **2013**, *4* (1), 1474.

(5) Kumar, K. S.; Dash, A. K.; H, H. S.; Verma, M.; Kumar, V.; Watanabe, K.; Taniguchi, T.; Gautam, G. S.; Singh, A. Towards a Comprehensive Understanding of the Low Energy Luminescence Peak in 2D Materials. arXiv December 22, 2023.

(6) Geim, A. K.; Grigorieva, I. V. Van Der Waals Heterostructures. *Nature* **2013**, *499* (7459), 419–425.

(7)  Tran, K.; Choi, J.; Singh, A. Moiré and beyond in Transition Metal Dichalcogenide Twisted Bilayers. *2D Mater.* **2020**, *8* (2), 022002.

(8)  Liu, Y.; Weiss, N. O.; Duan, X.; Cheng, H.-C.; Huang, Y.; Duan, X. Van Der Waals Heterostructures and Devices. *Nat Rev Mater* **2016**, *1* (9), 1–17.

(9)  Castellanos-Gomez, A.; Others. Deterministic Transfer of Two-Dimensional Materials by All-Dry Viscoelastic Stamping. *2D Mater.* **2014**, *1*.

(10) Pizzocchero, F.; Gammelgaard, L.; Jessen, B. S.; Caridad, J. M.; Wang, L.; Hone, J.; Bøggild, P.; Booth, T. J. The Hot Pick-up Technique for Batch Assembly of van Der Waals Heterostructures. *Nat Commun* **2016**, *7* (1), 11894.

(11) Nayak, P. K.; Horbatenko, Y.; Ahn, S.; Kim, G.; Lee, J.-U.; Ma, K. Y.; Jang, A.-R.; Lim, H.; Kim, D.; Ryu, S.; Cheong, H.; Park, N.; Shin, H. S. Probing Evolution of Twist-Angle-Dependent Interlayer Excitons in MoSe2/WSe2 van Der Waals Heterostructures. *ACS Nano* **2017**, *11* (4), 4041–4050.

(12) Kretinin, A. V.; Cao, Y.; Tu, J. S.; Yu, G. L.; Jalil, R.; Novoselov, K. S.; Haigh, S. J.; Gholinia, A.; Mishchenko, A.; Lozada, M.; Georgiou, T.; Woods, C. R.; Withers, F.; Blake, P.; Eda, G.; Wirsig, A.; Hucho, C.; Watanabe, K.; Taniguchi, T.; Geim, A. K.; Gorbachev, R. V. Electronic Properties of Graphene Encapsulated with Different Two-Dimensional Atomic Crystals. *Nano Lett.* **2014**, *14* (6), 3270–3276.

(13) Sanchez, D. A.; Dai, Z.; Wang, P.; Cantu-Chavez, A.; Brennan, C. J.; Huang, R.; Lu, N. Mechanics of Spontaneously Formed Nanoblisters Trapped by Transferred 2D Crystals. *Proceedings of the National Academy of Sciences* **2018**, *115* (31), 7884–7889.

(14) Liu, Y.; Liu, C.; Ma, Z.; Zheng, G.; Ma, Y.; Sheng, Z. Annealing Effect on Photoluminescence of Two Dimensional WSe2/BN Heterostructure. *Applied Physics Letters* **2020**, *117* (23), 233103.

(15) Vasu, K. S.; Prestat, E.; Abraham, J.; Dix, J.; Kashtiban, R. J.; Beheshtian, J.; Sloan, J.; Carbone, P.; Neek-Amal, M.; Haigh, S. J.; Geim, A. K.; Nair, R. R. Van Der Waals Pressure and Its Effect on Trapped Interlayer Molecules. *Nat Commun* **2016**, *7* (1), 12168.

(16) Shepard, G. D.; Ajayi, O. A.; Li, X.; Zhu, X.-Y.; Hone, J.; Strauf, S. Nanobubble Induced Formation of Quantum Emitters in Monolayer Semiconductors. *2D Mater.* **2017**, *4* (2), 021019.

(17) Radatović, B.; Çakıroğlu, O.; Jadriško, V.; Frisenda, R.; Senkić, A.; Vujičić, N.; Kralj, M.; Petrović, M.; Castellanos-Gomez, A. Strain-Enhanced Large-Area Monolayer MoS2 Photodetectors. *ACS Appl. Mater. Interfaces* **2024**, *16* (12), 15596–15604.

(18) Lee, K.-Y.; Lee, T.; Yoon, Y.-G.; Lee, Y.-J.; Cho, C.-H.; Rho, H. Raman Imaging of Strained Bubbles and Their Effects on Charge Doping in Monolayer WS2 Encapsulated with Hexagonal Boron Nitride. *Applied Surface Science* **2022**, *604*, 154489.

(19) Shabani, S.; Darlington, T. P.; Gordon, C.; Wu, W.; Yanev, E.; Hone, J.; Zhu, X.; Dreyer, C. E.; Schuck, P. J.; Pasupathy, A. N. Ultralocalized Optoelectronic Properties of Nanobubbles in 2D Semiconductors. *Nano Lett.* **2022**, *22* (18), 7401–7407.

(20) Tan, B. H.; Zhang, J.; Jin, J.; Ooi, C. H.; He, Y.; Zhou, R.; Ostrikov, K.; Nguyen, N.-T.; An, H. Direct Measurement of the Contents, Thickness, and Internal Pressure of Molybdenum Disulfide Nanoblisters. *Nano Lett.* **2020**, *20* (5), 3478–3484.

(21) Pan, Y.; Zahn, D. R. T. Raman Fingerprint of Interlayer Coupling in 2D TMDCs. *Nanomaterials* **2022**, *12* (22), 3949.

(22) Lim, S. Y.; Kim, H.; Choi, Y. W.; Taniguchi, T.; Watanabe, K.; Choi, H. J.; Cheong, H. Modulation of Phonons and Excitons Due to Moiré Potentials in Twisted Bilayer of $WSe_2$/$MoSe_2$. *ACS Nano* **2023**, acsnano.3c03883.

(23) Purdie, D. G.; Pugno, N. M.; Taniguchi, T.; Watanabe, K.; Ferrari, A. C.; Lombardo, A. Cleaning Interfaces in Layered Materials Heterostructures. *Nat Commun* **2018**, *9* (1), 5387.

(24) Zhang, B.; Zhang, Z.; Han, H.; Ling, H.; Zhang, X.; Wang, Y.; Wang, Q.; Li, H.; Zhang, Y.; Zhang, J.; Song, A. A Universal Approach to Determine the Atomic Layer Numbers in Two-Dimensional Materials Using Dark-Field Optical Contrast. *Nano Lett.* **2023**, *23* (19), 9170–9177.

(25) Li, H.; Wu, J.; Huang, X.; Lu, G.; Yang, J.; Lu, X.; Xiong, Q.; Zhang, H. Rapid and Reliable Thickness Identification of Two-Dimensional Nanosheets Using Optical Microscopy. *ACS Nano* **2013**, *7* (11), 10344–10353.

(26) Mondal, M.; Dash, A. K.; Singh, A. Optical Microscope Based Universal Parameter for Identifying Layer Number in Two-Dimensional Materials. *ACS Nano* **2022**, *16* (9), 14456–14462.

(27) Jeff Schewe, author. *The Digital Negative: Raw Image Processing in Lightroom, Camera Raw, and Photoshop / Jeff Schewe.*, Second edition.; Peachpit Press, 2016: San Francisco, 2016.

(28) Andrews, P.; Butler, Y.; Butler, Y. J.; Farace, J. *Raw Workflow from Capture to Archives: A Complete Digital Photographer's Guide to Raw Imaging*, 1st ed.; Focal Press, 2006.

(29) Dash, A. K.; Swaminathan, H.; Berger, E.; Mondal, M.; Lehenkari, T.; Prasad, P. R.; Watanabe, K.; Taniguchi, T.; Komsa, H.-P.; Singh, A. Evidence of Defect Formation in Monolayer MoS2 at Ultralow Accelerating Voltage Electron Irradiation. *2D Mater.* **2023**, *10* (3), 035002.

(30) Troparevsky, M. C.; Sabau, A. S.; Lupini, A. R.; Zhang, Z. Transfer-Matrix Formalism for the Calculation of Optical Response in Multilayer Systems: From Coherent to Incoherent Interference. *Opt. Express, OE* **2010**, *18* (24), 24715–24721.

(31) Anantharaman, S. B.; Stevens, C. E.; Lynch, J.; Song, B.; Hou, J.; Zhang, H.; Jo, K.; Kumar, P.; Blancon, J.-C.; Mohite, A. D.; Hendrickson, J. R.; Jariwala, D. Self-Hybridized Polaritonic Emission from Layered Perovskites. *Nano Lett.* **2021**, *21* (14), 6245–6252.

(32) Tran, K.; Moody, G.; Wu, F.; Lu, X.; Choi, J.; Kim, K.; Rai, A.; Sanchez, D. A.; Quan, J.; Singh, A.; Embley, J.; Zepeda, A.; Campbell, M.; Autry, T.; Taniguchi, T.; Watanabe, K.; Lu, N.; Banerjee, S. K.; Silverman, K. L.; Kim, S.; Tutuc, E.; Yang, L.; MacDonald, A. H.; Li, X. Evidence for Moiré Excitons in van Der Waals Heterostructures. *Nature* **2019**, *567* (7746), 71–75.

(33) Wang, K.; Huang, B.; Tian, M.; Ceballos, F.; Lin, M.-W.; Mahjouri-Samani, M.; Boulesbaa, A.; Puretzky, A. A.; Rouleau, C. M.; Yoon, M.; Zhao, H.; Xiao, K.; Duscher, G.; Geohegan, D. B. Interlayer Coupling in Twisted WSe2/WS2 Bilayer Heterostructures Revealed by Optical Spectroscopy. *ACS Nano* **2016**, *10* (7), 6612–6622.

(34) Zhang, X.; Qiu, J.; Qiu, J.; Qiu, J.; Li, X.; Li, X.; Zhao, J.; Zhao, J.; Liu, L.; Liu, L. Complex Refractive Indices Measurements of Polymers in Visible and Near-Infrared Bands. *Appl. Opt., AO* **2020**, *59* (8), 2337–2344.

(35) Haigh, S. J.; Gholinia, A.; Jalil, R.; Romani, S.; Britnell, L.; Elias, D. C.; Novoselov, K. S.; Ponomarenko, L. A.; Geim, A. K.; Gorbachev, R. Cross-Sectional Imaging of Individual Layers and Buried Interfaces of Graphene-Based Heterostructures and Superlattices. *Nature Mater* **2012**, *11* (9), 764–767.

(36) Ghorbanfekr-Kalashami, H.; Vasu, K. S.; Nair, R. R.; Peeters, F. M.; Neek-Amal, M. Dependence of the Shape of Graphene Nanobubbles on Trapped Substance. *Nat Commun* **2017**, *8* (1), 15844.

(37) Cao, P.; Xu, K.; Varghese, J. O.; Heath, J. R. The Microscopic Structure of Adsorbed Water on Hydrophobic Surfaces under Ambient Conditions. *Nano Lett.* **2011**, *11* (12), 5581–5586.

**Supporting information for**

# Quantification of 2D Interfaces: Quality of heterostructures, and what is inside a nanobubble

Mainak Mondal[1], Pawni Manchanda[2], Soumadeep Saha[1], Abhishek Jangid[1], Akshay Singh[1] *

[1] Department of Physics, Indian Institute of Science, Bengaluru, India- 560012

[2] Department of Physical Sciences, Indian Institute of Science Education and Research (IISER) Kolkata, Mohanpur, India- 741246

*Corresponding author: aksy@iisc.ac.in

Section 1: Transfer matrix model and ratio values calculations.

Section 2: RGB optical image and RAW format-based ratio maps for $MoS_2/MoS_2$ HS.

Section 3: AFM maps of $MoS_2/MoS_2$ HS before and after annealing.

Section 4: Photoluminescence map of $MoS_2/MoS_2$ HS after annealing.

Section 5: RGB optical image and RAW format-based ratio maps for $MoSe_2/WSe_2$ BL-HS.

Section 6: Variation of ratio values distribution width with increasing illumination.

Section 7: Quality index calculations for $MoS_2/MoS_2$ HS.

**Section 1: Transfer matrix model and ratio values calculations.**

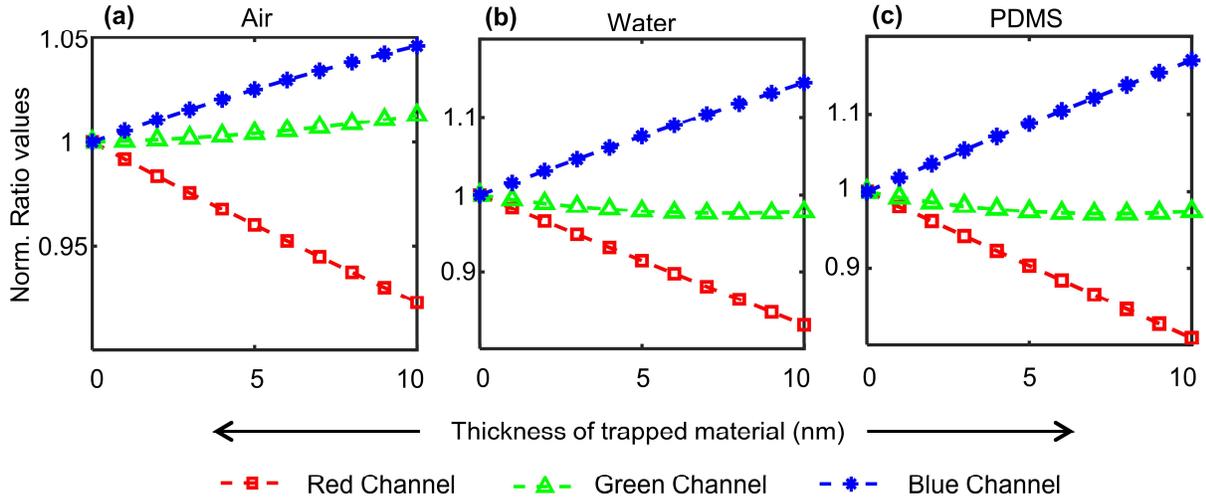

**Figure S1:** Calculated normalized ratio values for bilayer $MoS_2$/trapped material/monolayer $MoS_2$ on 257nm $SiO_2$/Si substrate system with changing trapped material thickness from 0 to 10 nm. The ratio values are normalized to the heterostructure with 0 nm trapped material thickness. **(a), (b) and (c)** show the calculations for air, water and PDMS, respectively.

**Section 2: RGB optical image and RAW format-based ratio maps for $MoS_2/MoS_2$ HS**

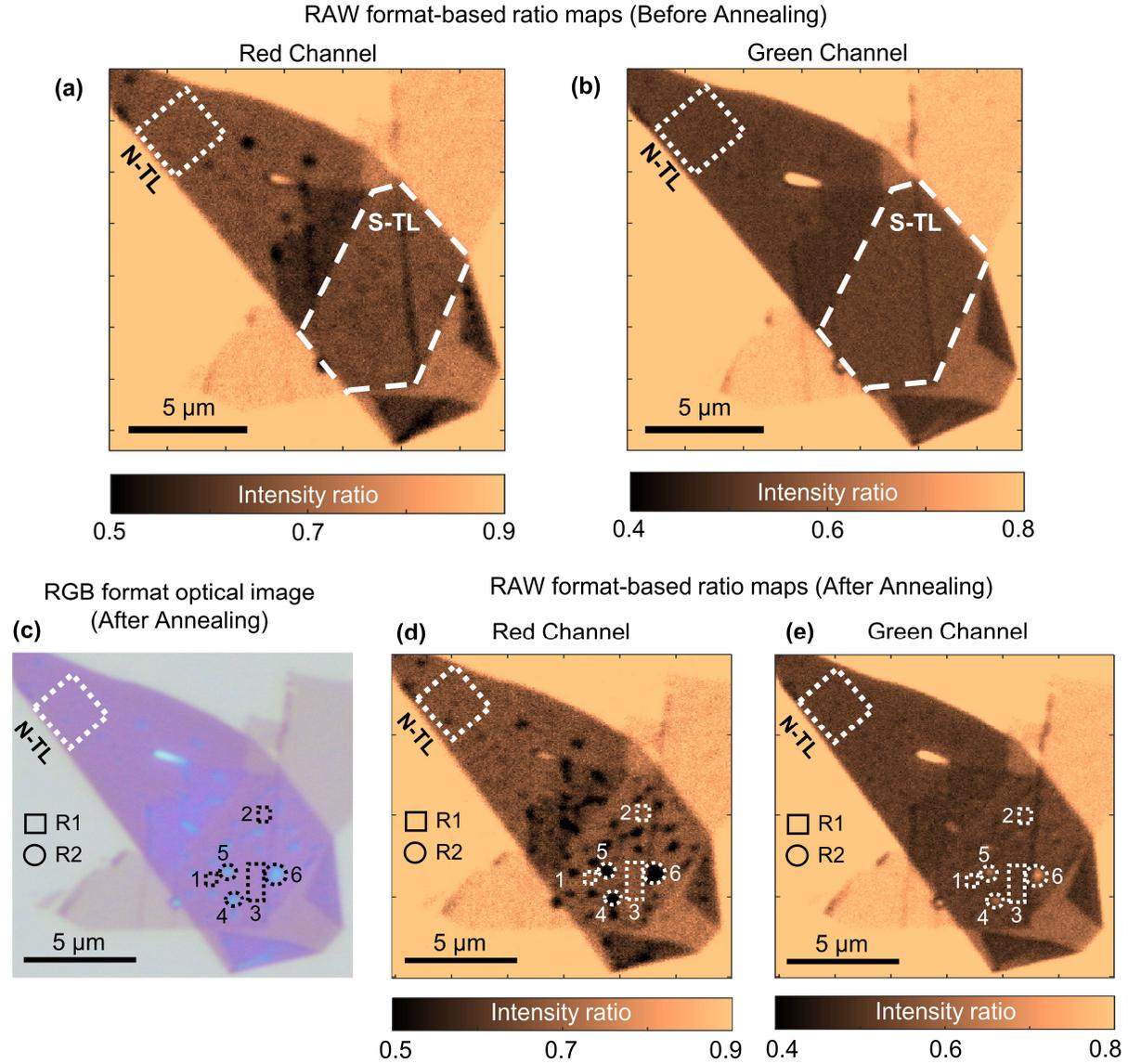

**Figure S2: (a) and (b)** RAW format-based red and green channel intensity ratio maps of the HS before annealing. Dashed lines on the image mark natural trilayer (N-TL) and stacked trilayer (S-TL) regions. $SiO_2/Si$ substrate with 257 nm $SiO_2$ thickness is used for this HS. **(c)** RGB optical image of the HS after annealing. **(d) and (e)** RAW format-based red and green channel intensity ratio maps of the HS after annealing. Regions appearing similar to (different from) N-TL are marked with rectangular (circular) apertures and named R1 (R2). $SiO_2/Si$ substrate with 257 nm $SiO_2$ thickness is used for this HS.

**Section 3: AFM maps of MoS$_2$/MoS$_2$ HS before and after annealing.**

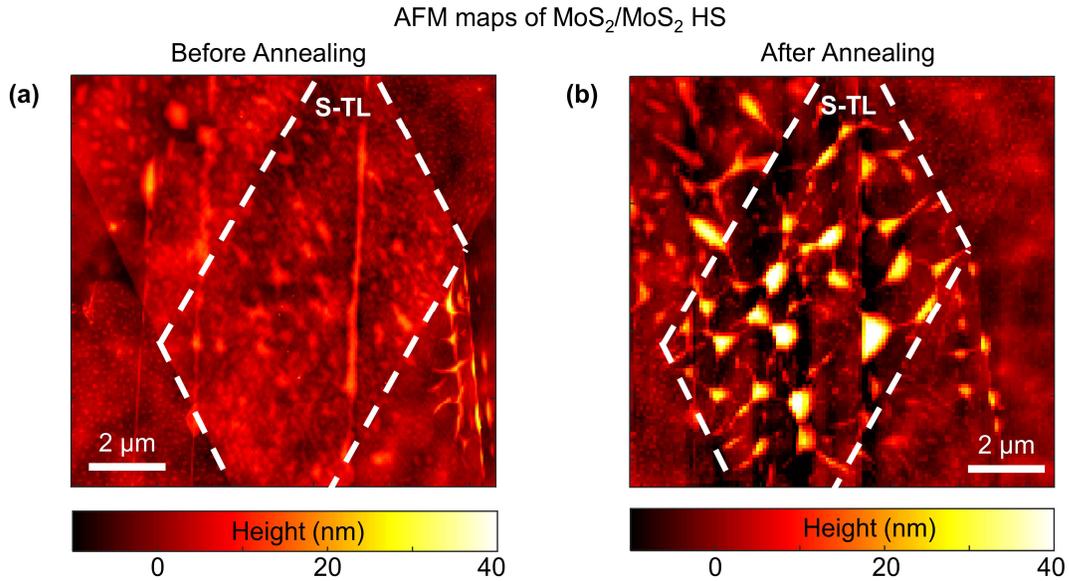

**Figure S3: (a) and (b)** Atomic Force Microscope (AFM) maps of the MoS$_2$/MoS$_2$ HS before and after annealing with the same height scale. Dashed lines on the figures map the S-TL region. The interface is very uniform before annealing as observed in the ratio maps. After annealing multiple nanobubbles are formed with height > 30 nm which we have categorized as R2 in main text figure 1 (c).

**Section 4: Photoluminescence map of MoS$_2$/MoS$_2$ HS after annealing.**

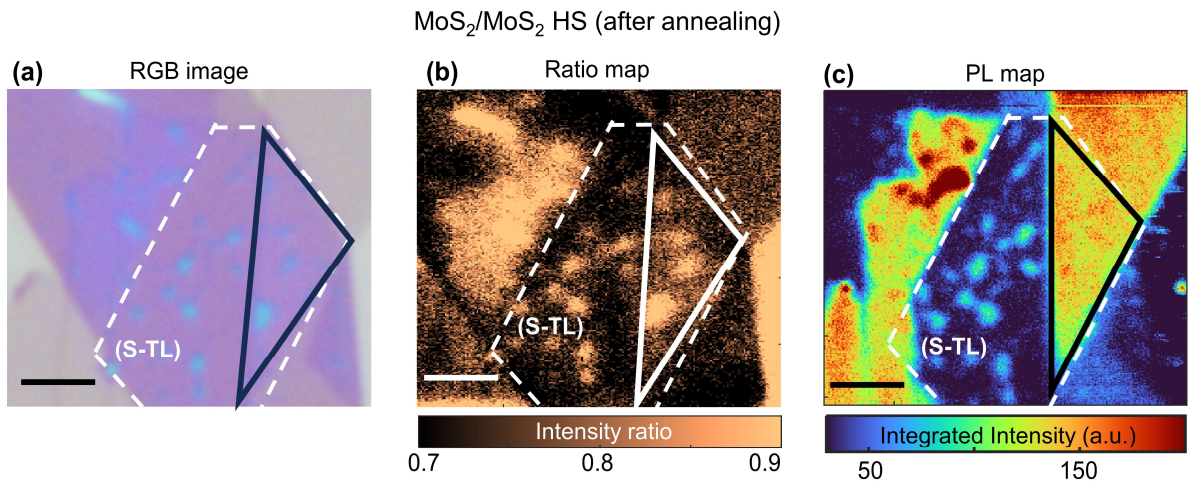

**Figure S4: (a), (b) and (c)** RGB-format image, blue channel ratio map and integrated PL map of the of $MoS_2/MoS_2$ HS region (after annealing), respectively. Dashed lines on the figures map the S-TL region. Marked triangular region represent the completely non-ideal region confirmed by the RAW ratio map and PL map. All scale bars are 2 μm.

**Section 5: RGB optical image and RAW format-based ratio maps for $MoSe_2/WSe_2$ BL-HS.**

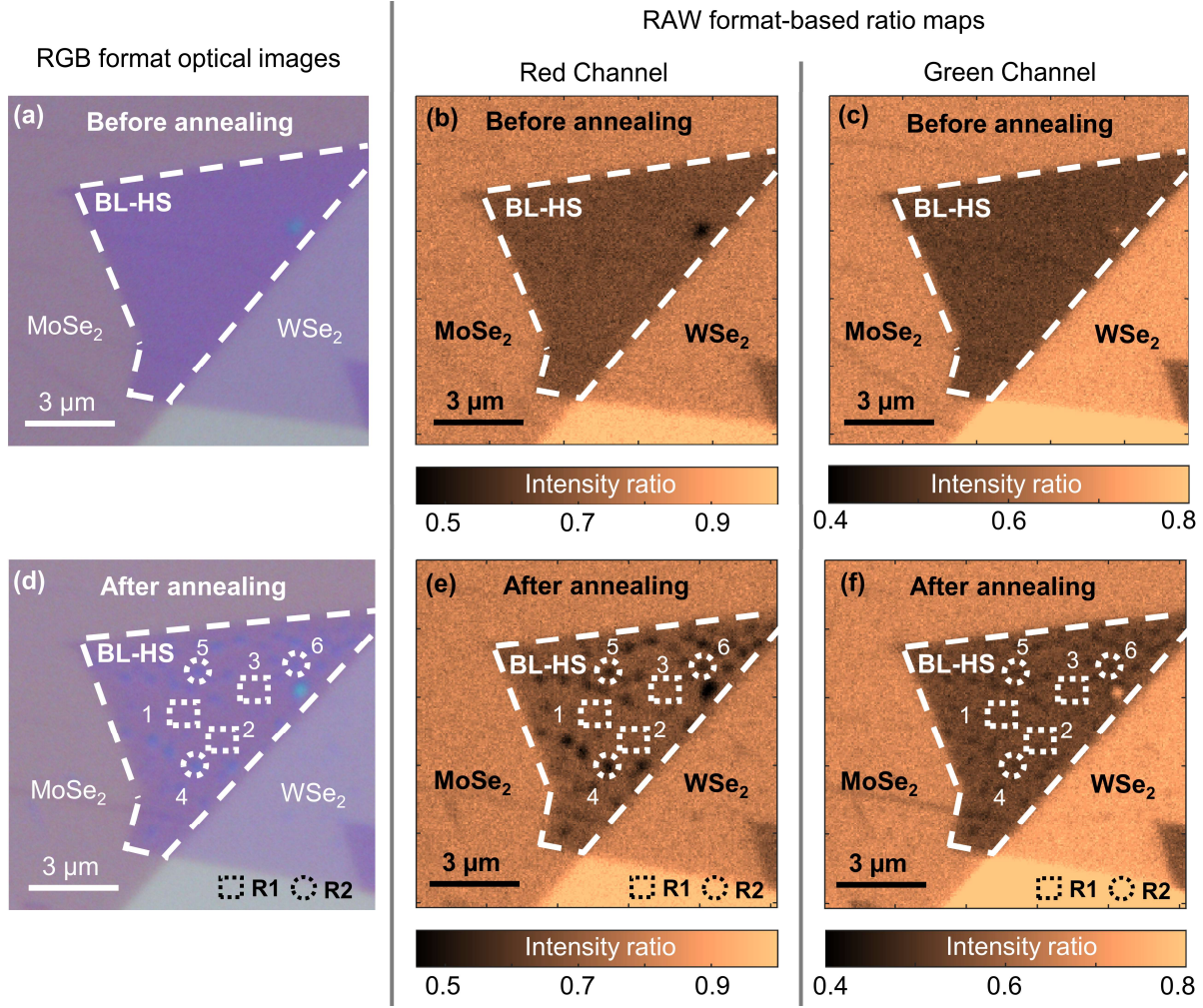

**Figure S5: (a)** RGB format image of the $MoSe_2/WSe_2$ BL-HS before annealing. **(b) and (c)** RAW format-based red and green channel ratio maps of the BL-HS before annealing. BL-HS region is marked with white boundaries, and individual layered regions are specified. **(d)** RGB format optical image of the $MoSe_2/WSe_2$ BL-HS after annealing. **(e) and (f)** RAW format-based red and green channel ratio maps of the BL-HS after annealing. BL-HS region is marked with white boundaries, and individual layered regions are specified on all the figures. For after-annealing image and maps, regions where the ratio values have remained the same (decreased) after annealing are marked with rectangular (circular) apertures and categorized as R1 (R2). $SiO_2/Si$ substrate with 257nm $SiO_2$ thickness is used for this HS.

**Section 6: Variation of ratio values distribution width with increasing illumination.**

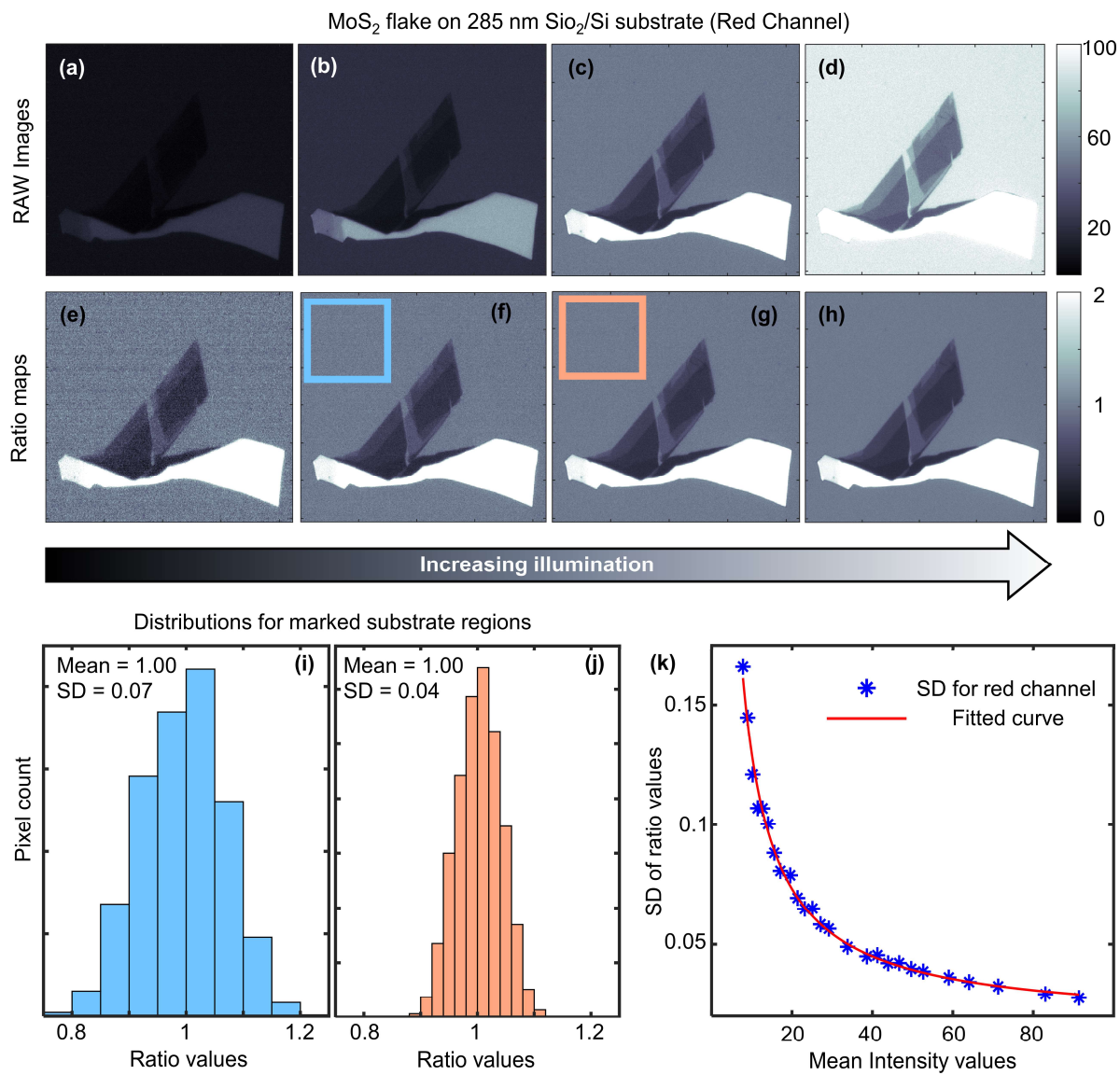

**Figure S6: (a) to (d)** Extracted RAW-format red channel images with increasing imaging lamp illumination of a $MoS_2$ flake on a 285nm $SiO_2/Si$ substrate. Colorbar represents the pixels' intensity, same for all images **(e) to (f)** Red channel ratio maps calculated from the images shown in figure (a) to (d). Colorbar represents

pixels' ratio values, same for all the maps. **(i) and (j)** Distribution of the number of pixels with ratio values taken from the substrate region (marked in figure (f) and (g)), respectively. Mean and standard deviation (SD) of the distribution is mentioned in figure labels. **(k)** Variation of SD values with increasing mean intensity values is shown as star markers and red solid line is fitted curve. All shown results are for the red channel.

An exfoliated $MoS_2$ flake on a 285nm $SiO_2$/Si substrate is imaged in RAW-format with increasing imaging lamp illumination. The extracted red channel images are shown from figure (a) to (d) with increasing illumination order. Corresponding ratio maps of these images are shown from figure (e) to (h) in the same order. It is evident from the ratio maps that there was a significant amount of noise present in the images for low illumination, and noise reduces with increasing illumination. The distribution of ratio values from a section of the substrate (marked in a blue box in figure (f) and (g)) is shown in figure (i) and figure (j). Corresponding mean and standard deviation (SD) of this distribution is mentioned in figure labels. The SD of this distribution is the direct effect of noise present in the CMOS imaging sensor for a certain illumination. With increasing intensity, such as from figure (f) to figure (g), the SD has reduced almost by 50%.

Next, we have plotted the noise as SD from each ratio map with the corresponding mean intensity value from the RAW images. Please note that for RAW-format, the mean intensity values increase linearly with illumination, and hence represents the illumination intensity for that image. The noise has a clear decrease with increasing illumination, and it can be fitted with $y = a/x + b$ equation. For this fitting $a = 1.13$ and $b = 0.017$. Hence, even at maximum intensity (255 for our camera) the minimum SD will be 0.02. Green and blue channel shows the same variation as the red channel. In our measurement shown in the main text the SD is 0.03, which indicates excellent signal to noise. The noise can be further reduced by averaging multiple frames which we have not performed in this study.

## Section 7: Quality index calculations for MoS$_2$/MoS$_2$ HS.

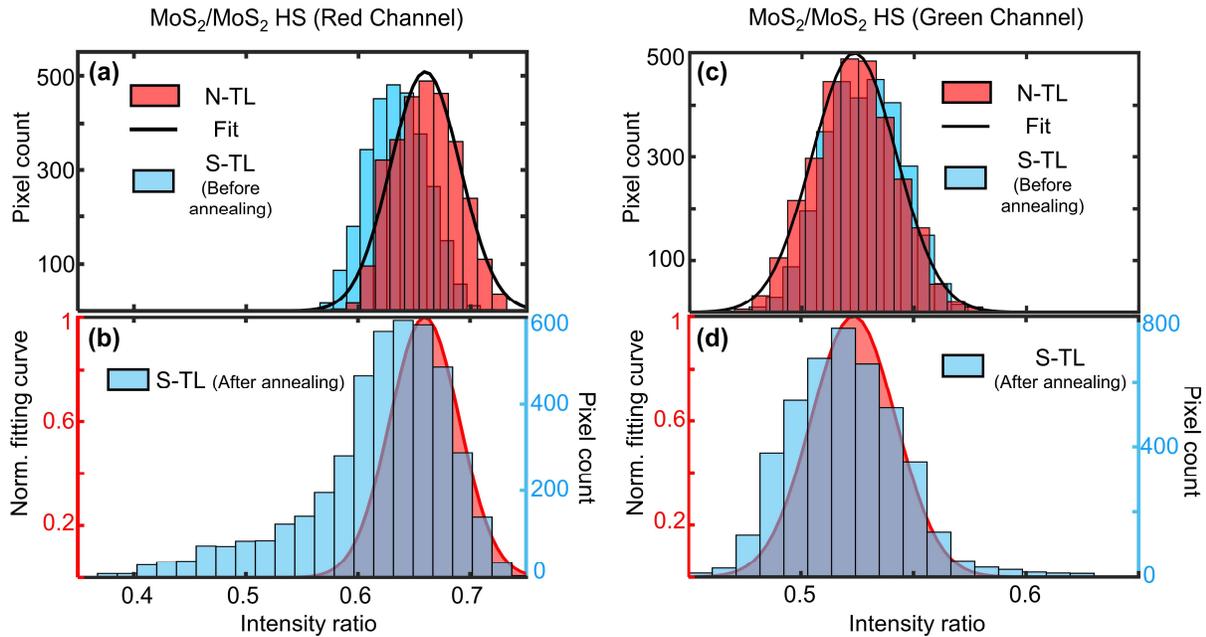

**Figure S7: (a)** Histogram showing the distribution of pixel counts with ratio values for N-TL and S-TL before annealing. Gaussian fit of the N-TL distribution is shown in a solid black line. **(b)** The ratio values distribution for the after-annealing S-TL region (shown in main text figure 3(b) inset) is plotted on the right y-axis. The Y-axis left shows the normalized fitted curve for the N-TL region. This data is for the MoS$_2$/MoS$_2$ device, red channel. Same analysis for green channel is shown in figure **(c)** and **(d)**. Quality index for the red and green channel is 0.48 and 0.62 respectively.